\documentclass[10pt,journal,comsoc]{IEEEtran}

\usepackage{cite}
\usepackage{amsmath,amssymb,amsfonts}
\usepackage[normalem]{ulem}
\usepackage{graphicx}
\usepackage{textcomp}
\usepackage{svg}      
\usepackage{xcolor}
\usepackage[most]{tcolorbox}
\usepackage[table]{xcolor}
\usepackage[caption=false,font=footnotesize]{subfig} 
\usepackage[
    hidelinks,
    bookmarks=false
]{hyperref}
\usepackage{listings}
\usepackage{multirow}
\usepackage{graphicx}     
\tcbset{colback=gray!10!white, colframe=gray!30!black, boxrule=0.4pt, arc=2pt,
        left=6pt,right=6pt,top=6pt,bottom=6pt, fonttitle=\bfseries}
\usepackage{comment}
\usepackage{pifont}
\usepackage{booktabs}
\usepackage{colortbl}
\usepackage{array}

\tcbset{
  colback=white, colframe=black!35,
  boxsep=1pt, left=4pt, right=4pt, top=3pt, bottom=3pt,
  before skip=4pt, after skip=4pt,
  fonttitle=\bfseries\footnotesize
}
\usepackage{algpseudocode}\usepackage[linesnumbered, ruled]{algorithm2e}
\usepackage[linesnumbered, ruled]{algorithm2e}

\SetKwRepeat{Do}{do}{while}%

\SetCommentSty{mycommfont}

\makeatletter
\newcommand{\algmargin}{\the\ALG@thistlm}
\makeatother

\newlength{\forwidth}
\algdef{SE}[parFOR]{parFor}{EndparFor}[1]
{\parbox[t]{\dimexpr\linewidth-\algmargin}{%
\hangindent\forwidth\strut\algorithmicfor\ #1\ \algorithmicdo\strut}}{\algorithmicend\ \algorithmicfor}%

\newlength{\whilewidth}
\algdef{SE}[parWHILE]{parWhile}{EndparWhile}[1]
{\parbox[t]{\dimexpr\linewidth-\algmargin}{%
\hangindent\whilewidth\strut\algorithmicwhile\ #1\ \algorithmicdo\strut}}{\algorithmicend\ \algorithmicwhile}%
\algnewcommand{\parState}[1]{ %
\parbox[t]{\dimexpr\linewidth-\algmargin}{\strut #1\strut}}

\usepackage{epstopdf}

\begin{document}

\title{Intent2Tc: Automated Intent-to-Traffic Control Translation with Language Models}
\author{
\IEEEauthorblockN{Andrea Masini, Sudipta Acharya,~Paolo Bellavista,~Luca Foschini,~Burak Kantarci \vspace{-0.2in}}\\
\thanks{Andrea Masini, Sudipta Acharya and Burak Kantarci are with the University of Ottawa, Ottawa, ON, Canada. Emails: \{amasi092,sacharya2,burak.kantarci\}@uottawa.ca\\
Andrea Masini, Paolo Bellavista and Luca Foschini are with the University of Bologna, Italy. Emails: andrea.masini6@studio.unibo.it \& \{paolo.bellavista,luca.foschini\}@unibo.it\\
This work was performed while Andrea Masini was affiliated with the University of Ottawa, working in B. Kantarci's research lab. 
}
}


\IEEEtitleabstractindextext{
\begin{abstract}
Automated and highly usable Quality-of-Service (QoS) enforcement requires translating high-level service intents into deployable traffic-management policies. Although intent-based networking (IBN) has simplified policy specification, bridging the gap between business-level intents and executable network configurations remains complex, error-prone, and difficult to automate. This paper presents Intent2Tc, a closed-loop language-model-driven framework that translates business-level traffic-shaping intents into declarative sub-intents and subsequently into validated, executable Linux traffic control (\texttt{tc}) configurations. The framework integrates an Active Queue Management (AQM)-based digital twin (DT) semantic model, automated metadata extraction, critique-driven refinement, and Retrieval-Augmented Generation (RAG)-based knowledge reuse to improve semantic consistency and configuration reliability. We evaluate multiple open-source large language models (LLMs) and small language models (SLMs), together with Claude Sonnet-4.6, on 100 Request for Comments (RFC) 9315-compliant traffic-shaping intents. Across both translation stages, Intent2Tc achieves high semantic fidelity, configuration accuracy, and deployment readiness, with Claude Sonnet-4.6 reaching 0.98 semantic similarity, 1.0 semantic unit coverage, and 0.045 normalized edit distance. Furthermore, RAG reduces token consumption and inference latency while enabling compact models such as Phi-4-mini to approach the performance of substantially larger models. Linux \texttt{tc} serves as the target configuration platform, demonstrating the practical applicability of the proposed framework.
\end{abstract}

\begin{IEEEkeywords}
IBN, Linux \texttt{tc}, language models, RAG.
\end{IEEEkeywords}}
\pagestyle{empty}

\maketitle
\thispagestyle{empty}
\IEEEdisplaynontitleabstractindextext
\IEEEpeerreviewmaketitle

\section{Introduction}\label{intro}
Interest in IBN frameworks has grown significantly in recent years~\cite{leivadeas2022survey,clemm2022intent}, driven by the need to simplify traffic management and enable autonomous actuation, supervision, and adaptation across various scenarios~\cite{velasco2021end}. Research shows that most existing IBN work focuses on the underlying infrastructure layer~\cite{yu2023comprehensive}, while the intent-processing pipeline remains dependent on slow, repetitive, and error-prone manual creation of network-shaping rules. More recently, language models (LMs) have emerged as a pivotal approach for intent translation, mediating between high-level human articulations and the underlying network configuration primitives~\cite{dzeparoska2023llm, dinh2025towards}. Despite notable advances in the validation and orchestration stages~\cite{dinh2025towards}, the terminal translation phase is largely left to manual intervention~\cite{wang2024netconfeval}. Intent2Tc addresses this gap as a natural-language-to-\texttt{tc} translator focused on the acquisition-and-translation stage of the IBN lifecycle.

Recently, the authors of~\cite{acharya2026intent2qos} introduced the first queueing-theory-based end-to-end LM-driven intent-to-\texttt{tc} translation pipeline, but it relied on static keyword matching and a single-pass critique, which limited recovery from mismatches and learning from validated outputs. To overcome these limitations, we propose a \emph{closed-loop} framework with dynamic metadata extraction, multi-stage critique, and RAG-based knowledge reuse to improve accuracy, robustness, and adaptability.

Our framework follows a three-phase workflow: (1) an AQM-based DT provides a \textit{semantic model} of the simulated priority-queueing network; (2) an LM extracts metadata (aided by a confidence-based common-patterns lookup table (LUT)), decomposes the intent into declarative \textit{sub-intents}, and generates Linux \texttt{tc} rules, while a Critique Module (CM) corrects hallucinations, duplicates, and omissions after each stage; (3) the RAG database (RAG DB) is updated with the corrected intent, sub-intents, and \texttt{tc} rules, and indexed by traffic profile and time bounds. The key contributions of this work are as follows:
\begin{itemize}
\item A scalable DT environment implementing priority-class queueing with AQM, enabling semantic-model generation for diverse traffic-shaping scenarios.
\item A RAG DB of validated intents, sub-intents, and \texttt{tc} rules to support continual improvement in intent translation.
\item A closed-loop LM-driven framework for translating traffic-shaping intents into deployable \texttt{tc} rules through DT-based semantic modeling, dynamic traffic-profile matching, confidence-based metadata extraction, critique-driven correction, and RAG-assisted knowledge reuse.
\end{itemize}

To evaluate end-to-end intent-to-\texttt{tc} translation, we transform a publicly available business-intent dataset~\cite{li2025business} into 100 RFC 9315-compliant traffic-shaping intents~\cite{clemm2022intent}, preserving their real-world context. We evaluate multiple open-source LLMs and SLMs available on Hugging Face~\cite{huggingface} with and without RAG augmentation, including Hermes-2-Pro (8B), Qwen-3 (4B/8B), Phi-4-mini (3.8B), Llama-3.1 (8B), Qwen3.5-Claude-Opus4.6-Distill (9B), Gemma-4-E4B (8B), and IBM Granite-4.1 (8B), using Claude Sonnet-4.6 as a closed-source baseline. The reported results show highly accurate traffic-profile and time-bound extraction, while RAG reduces token usage and inference latency. Claude Sonnet-4.6 achieves 0.98 semantic similarity, 1.0 semantic unit coverage, and 0.045 normalized edit distance.

The remainder of this paper is structured as follows. Section~II reviews related work and identifies the gap in closed-loop intent-to-\texttt{tc} translation. Section~III presents the proposed framework, including semantic modeling, metadata extraction, critique-driven correction, and RAG-based knowledge reuse. Section~IV describes the dataset, evaluation metrics, case study, and results. Section~V concludes the paper and outlines future work.

\newcommand{\cmark}{\ding{51}} 
\newcommand{\xmark}{\ding{55}} 

\section{Related Work}\label{relatedWorks}
Linux \texttt{tc} is a widely used but complex tool for enforcing QoS in cloud-native environments such as Kubernetes. It supports classful and classless schedulers, packet filtering, priority-based scheduling, and traffic-shaping mechanisms, such as AQM. However, its strict syntax and kernel-level complexity make manual rule generation difficult and error-prone~\cite{pfefferle2021ieee}. Recent work shows that LMs can help parse natural-language intents~\cite{bimo2025intent} and automate parts of the intent lifecycle, but the final step of turning validated intents into platform-specific configurations remains challenging per RFC 9315~\cite{clemm2022intent}. Frameworks such as INTA~\cite{wei2025inta} use intents as an intermediate representation to bridge heterogeneous configuration models, while NetConfEval~\cite{wang2024netconfeval} shows that LLMs can translate high-level policies into formal specifications for software-defined networking (SDN) controllers; however, support for the more complex syntax of Linux \texttt{tc} remains limited. NetIntent~\cite{hossain2025netintent} automates the intent lifecycle using LLMs for conflict detection and resolution. However, most LLMs remain prone to hallucinations and can produce technically plausible but incorrect configurations.

Recently, the authors of~\cite{acharya2026intent2qos} attempted to close this gap, proposing an LM-based pipeline for intent-to-QoS rule generation. In their work, they assess the performance of different prompting strategies in sub-intent generation and \texttt{tc} rule translation, identifying few-shot prompting as the best-performing approach. They specifically state the need for an iterative learning loop, describing it as a key missing component of their infrastructure.

Overall, existing studies improve intent interpretation, validation, and configuration generation, but they mostly rely on a single-pass translation process. This leaves limited room for knowledge reuse, iterative refinement, or learning from previous translations, motivating a closed-loop intent-to-\texttt{tc} framework with semantic modeling, critique-based validation, and retrieval-driven knowledge accumulation.

\begin{figure}[!htb]
  \centering
\includegraphics[width=1\linewidth]{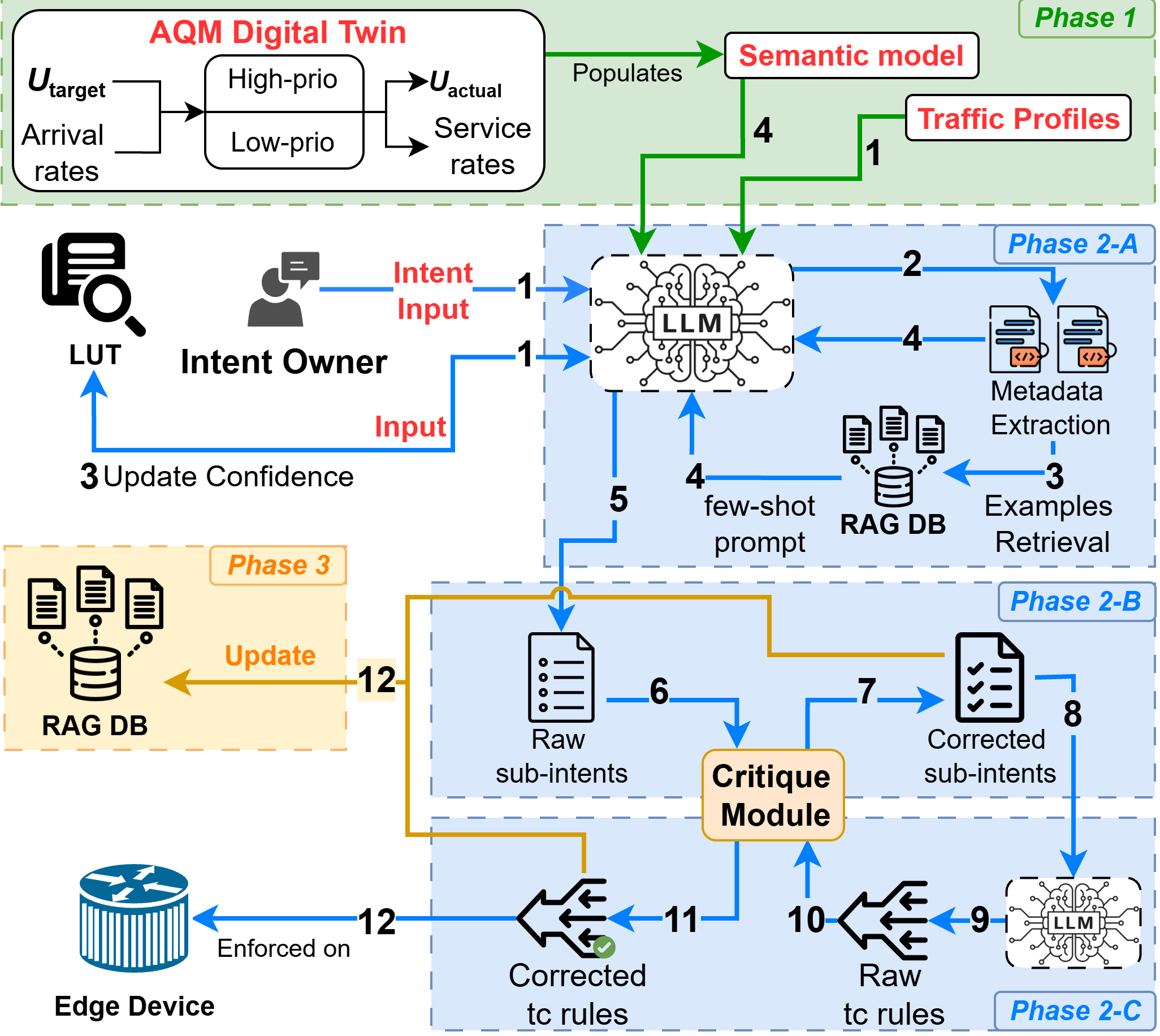}
  \caption{Proposed intent-to-\texttt{tc} pipeline: numbered arrows show execution order; unnumbered arrows show independent contextual flows.}
  
  \label{fig:intent-pipeline}
\end{figure}

\section{Methodology}
Fig.~\ref{fig:intent-pipeline} illustrates the proposed architecture and intent translation loop, while Algorithm~\ref{alg:intent2tc} summarizes the overall workflow.

\paragraph{Phase 1: AQM DT Semantic Model Setup}
The initial phase builds a \texttt{SimPy}-based network DT using a finite-capacity, non-preemptive queueing model with priority classes calibrated offline using queueing theory. It uses Controlled Delay (CoDel)-based AQM to favor dropping low-priority packets under overload, while remaining extensible to other mechanisms, such as Random Early Detection (RED). Traffic arrivals follow a Poisson process, and service rates are tuned to meet target utilization levels. During simulation, packets are scheduled according to priority and managed according to queue policies under overload conditions. The resulting performance metrics are organized into a semantic model that links traffic objectives to feasible system behavior and supports subsequent phases of intent interpretation and configuration generation.

\paragraph{Phase 2: Metadata Extraction, Intent Subdivision, and Rule Generation}
In the second phase, an LM-driven pipeline transforms high-level traffic-shaping intents into deployable Linux \texttt{tc} rules using four sources of context: the user intent, a confidence-based LUT, the semantic model from \emph{Phase 1}, and a traffic-profile taxonomy.

\emph{Phase 2-A -- Helper Metadata Extraction:}
The LM extracts auxiliary metadata from the user intent, including the traffic profile, priority level, and time bounds, to guide subsequent translation stages. The confidence-based LUT maps recurring natural-language expressions to priority and temporal constraints, providing additional contextual guidance.

When RAG is enabled, the system retrieves the two most relevant examples from the RAG DB that match the extracted traffic profile and injects them into the sub-intent and rule-generation prompts. Otherwise, it falls back to a generic few-shot prompting strategy. Each RAG DB entry stores a \textit{slot} (\texttt{traffic\_profile::time\_bounds}), the original \texttt{intent}, corrected \texttt{sub-intents}, corrected \texttt{tc rules}, and the associated \texttt{Token-F1} score.

\emph{Phase 2-B1 -- Generation of Declarative Sub-Intents:}
From the high-level intent, the LM generates declarative sub-intents expressing network requirements and queueing policies. Collectively, they form a fundamental intermediate interface bridging the general intent and low-level configuration. The prompt draws from three main context sources: extracted metadata, the semantic model, and either dynamic RAG-retrieved or static few-shot examples. The generated sub-intents capture performance constraints, such as delay and packet-drop thresholds; traffic prioritization requirements, including the appropriate priority class; AQM-related thresholds for congestion management; and temporal constraints governing policy enforcement.

\emph{Phase 2-B2 -- Sub-Intent Correction:}
The generated sub-intents are subsequently validated by a deterministic, template-based CM. The CM identifies and removes invalid or policy-inconsistent sub-intents, inserts any missing requirements implied by the original intent and semantic model, and merges semantically redundant entries. The result is a complete, consistent, and logically coherent set of corrected sub-intents for downstream rule generation.

\emph{Phase 2-C1 -- Low-Level Linux \texttt{tc} Rule Generation:}
In the final stage of Phase 2, the LM translates the corrected sub-intents into executable Linux \texttt{tc} rules. The generation process is guided by traffic-profile attributes (e.g., Internet Protocol (IP) addresses, ports, and protocols), the semantic model, and either RAG-retrieved or few-shot examples to ensure that the resulting configurations remain consistent with both the intent requirements and the underlying traffic characteristics.

\emph{Phase 2-C2 -- \texttt{tc} Rule Correction:}
The generated \texttt{tc} rules are then validated and refined by the CM to produce a consistent, syntactically correct, and deployment-ready configuration set.
The CM enforces key \texttt{tc}-specific constraints, including the Hierarchical Token Bucket (HTB) class hierarchy (e.g., \texttt{classid 1:10/1:11}), priority mapping (e.g., \texttt{prio 0/prio 2}), \texttt{fq\_codel} AQM parameters, \texttt{flower}-based packet filtering, and strict syntactic ordering, since swapping two operators in a \texttt{tc} rule can trigger cascading failures. All modifications are logged to provide an auditable trail of the translation and correction process.
\begin{algorithm}[!htp]
\caption{Intent-to-\texttt{tc} configuration: semantic modeling, critique correction, and RAG optimization}
\label{alg:intent2tc}
\fontsize{7.3}{7.5}\selectfont

\KwData{Natural-language intents $\mathcal{I}_{\text{list}}$, critique templates (sub-intents $R_d$, \texttt{tc} rules $R_c$), and RAG flag \texttt{USE\_RAG}}
\KwResult{Corrected sub-intents $\hat{I}$ and valid \texttt{tc} rules $T_{\hat{I}}$}

\textbf{/* Phase 1: Semantic Modeling */}\\
Set $(\lambda_{\text{high}}, \lambda_{\text{low}}, \mu_{\text{high}}, \mu_{\text{low}})$;
$S \leftarrow$ SimPy priority queue with CoDel\;

\textbf{/* Phase 2: Main Loop */}\\
Load common patterns $CP$, traffic profiles $TP_{\text{list}}$, and few-shot examples $E$\;

\ForEach{$i \in \mathcal{I}_{\text{list}}$}{
  \textbf{/* Phase 2-A: Metadata Extraction */}\\
  $M_d \leftarrow \text{LM\_extract}(\text{\emph{build\_ext\_prompt}}(i, CP, TP_{\text{list}}))$\;
  $(Prof, Prio, Time, Synt_{id}, Synt_{con}, Time_{id}, Time_{con}) \leftarrow M_d$\;
  \emph{update\_CP}$(Synt_{id}, Synt_{con}, Time_{id}, Time_{con})$\;
  $P \leftarrow \text{profile\_filter}(Prof)$\;
  $(R_d, R_c) \leftarrow \text{build\_critique}(M_d, P)$\;
  \If{\texttt{USE\_RAG}}{$E \leftarrow \emph{rag\_retrieve}(P::Time)$}

  \textbf{/* Phase 2-B: Sub-intent Generation */}\\
  $\hat{I} \leftarrow \text{fix\_subs}\!\left(\text{LM\_subintents}(\text{\emph{build\_sub\_prompt}}(i, S, M_d, E)), R_d\right)$\;

  \textbf{/* Phase 2-C: Linux \texttt{tc} Rules Generation */}\\
  $T_{\hat{I}} \leftarrow \text{fix\_tc}\!\left(\text{LM\_tc}(\text{\emph{build\_cfg\_prompt}}(\hat{I}, S, P, E)), R_c\right)$\;

  \textbf{/* Phase 3: RAG DB Update */}\\
  \If{\texttt{USE\_RAG}}{$\text{RAG DB} \leftarrow \emph{rag\_update}(P::Time, i, \hat{I}, T_{\hat{I}}, \text{compute\_TokenF1})$}

  \textbf{/* Output */}\\
  Save $\hat{I}$ and $T_{\hat{I}}$ as JavaScript Object Notation (JSON); log corrections and motivations\;
}
\end{algorithm}

\paragraph{Phase 3 -- Updating the RAG DB}
Finally, when RAG is enabled, the system evaluates the corrected translation for possible inclusion in the RAG DB. A \texttt{Token-F1} score is computed, and the new entry is inserted only if its score exceeds that of at least one existing record. 
Token-F1 was selected as a lightweight proxy for pre-deployment comparison, capturing token-level overlap between generated and ground-truth outputs.
The RAG DB is initially empty and is progressively populated with validated intent–sub-intent–\texttt{tc} rule mappings as the system processes successive intents.
\begin{figure}[!htp]
  \centering
  \subfloat[QoS-goal shares]{%
    \includegraphics[width=0.45\linewidth]{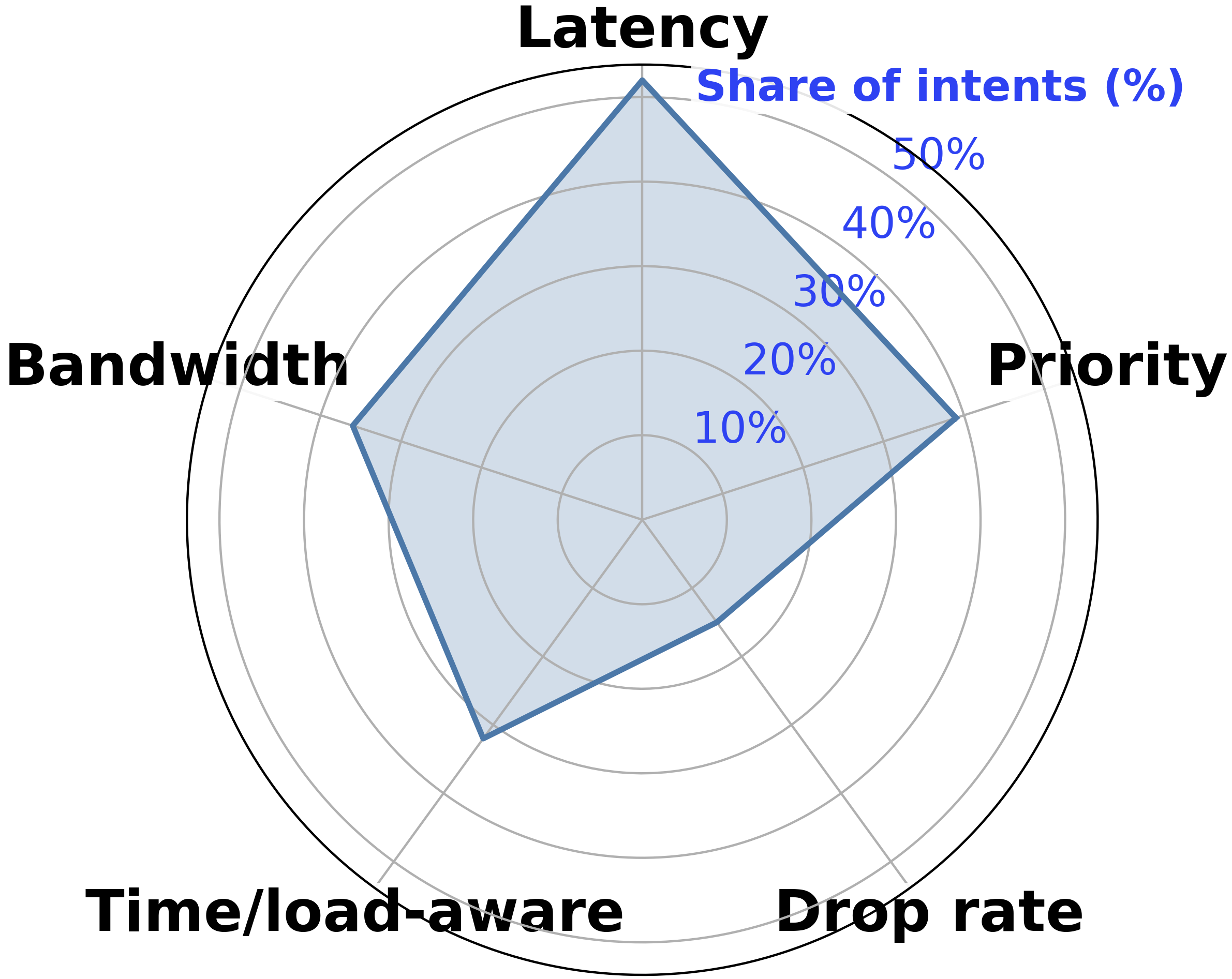}%
    \label{fig:qos}}
  \hfill
  \subfloat[Traffic-profile distribution]{%
    \includegraphics[width=0.40\linewidth]{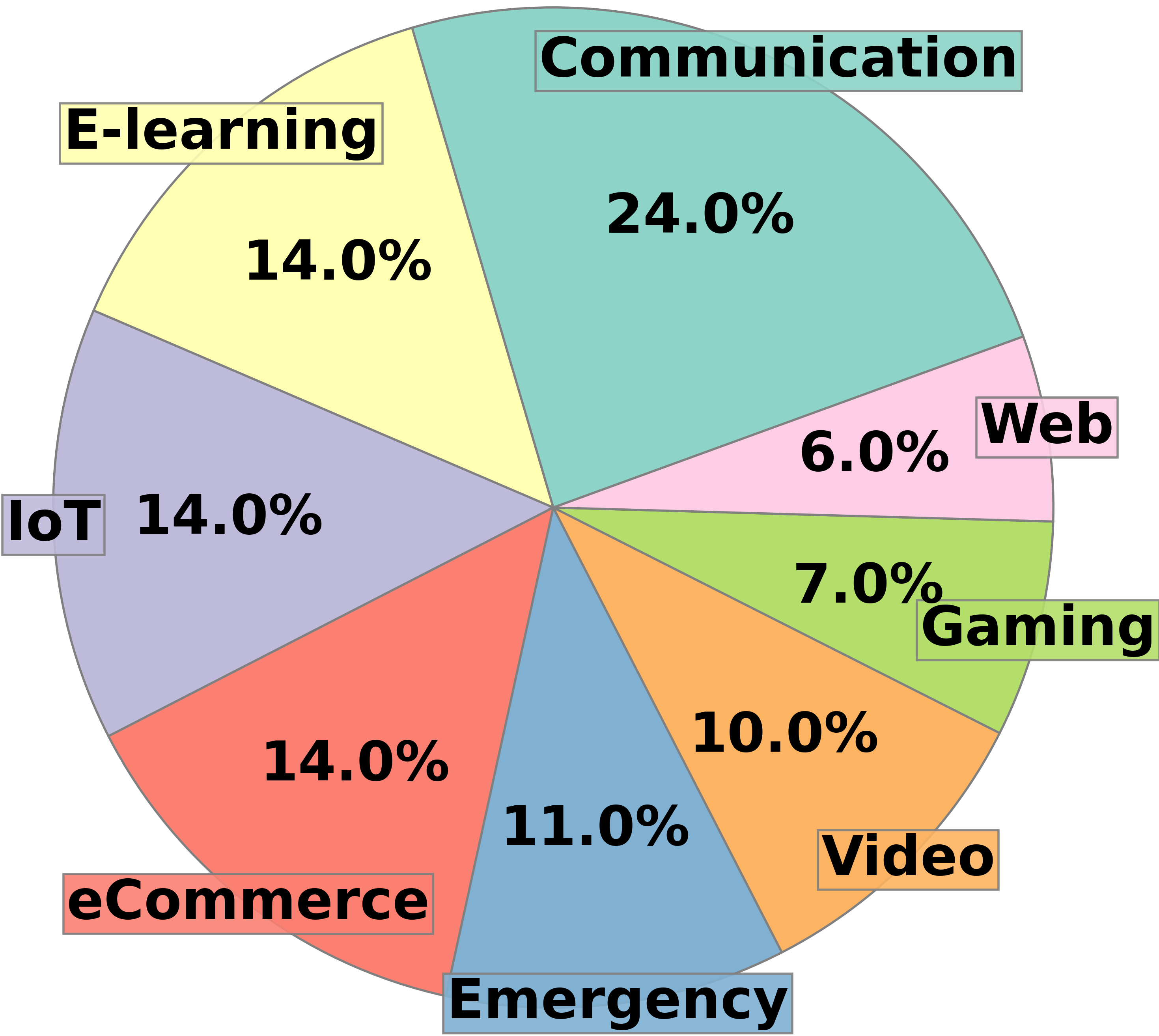}%
    \label{fig:traffic}}

  \vspace{0.05em}
  \subfloat[Intent-constraint structure]{%
    \includegraphics[width=0.70\linewidth]{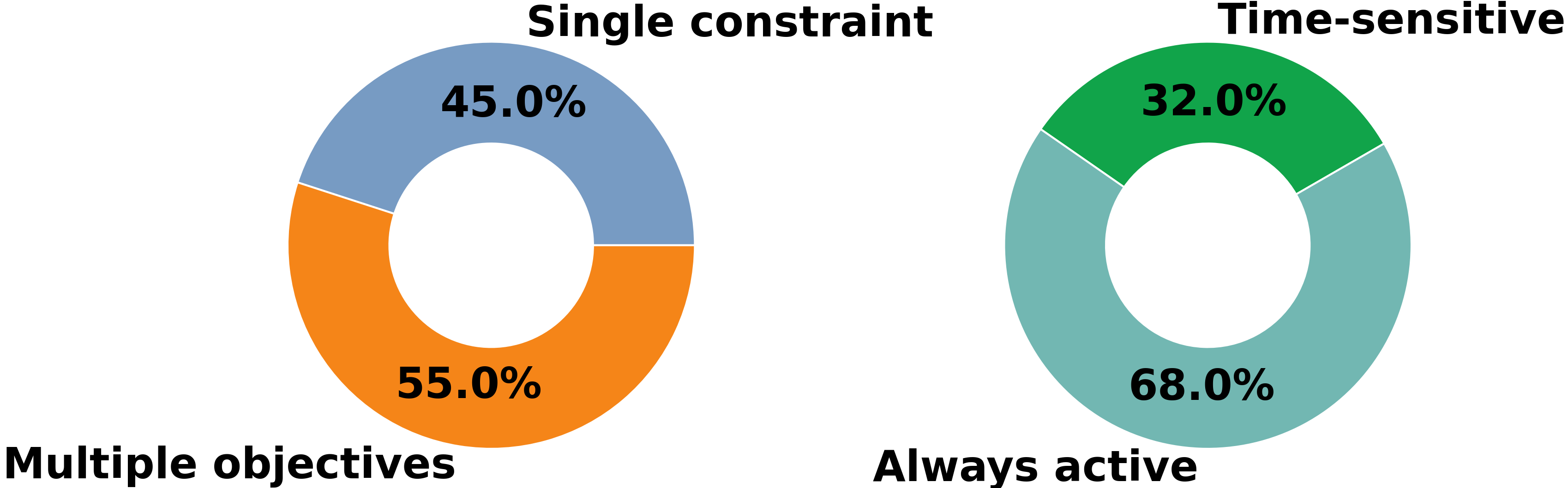}%
    \label{fig:intent}}

  \caption{100-intent dataset statistics} 
  \label{fig:dataset-stats}
\end{figure}

\paragraph{Output}
The final output of the framework is a validated and deployment-ready set of Linux \texttt{tc} rules for traffic-shaping enforcement. 

\begin{figure*}[t]
  \centering
\includegraphics[width=1.0\linewidth]{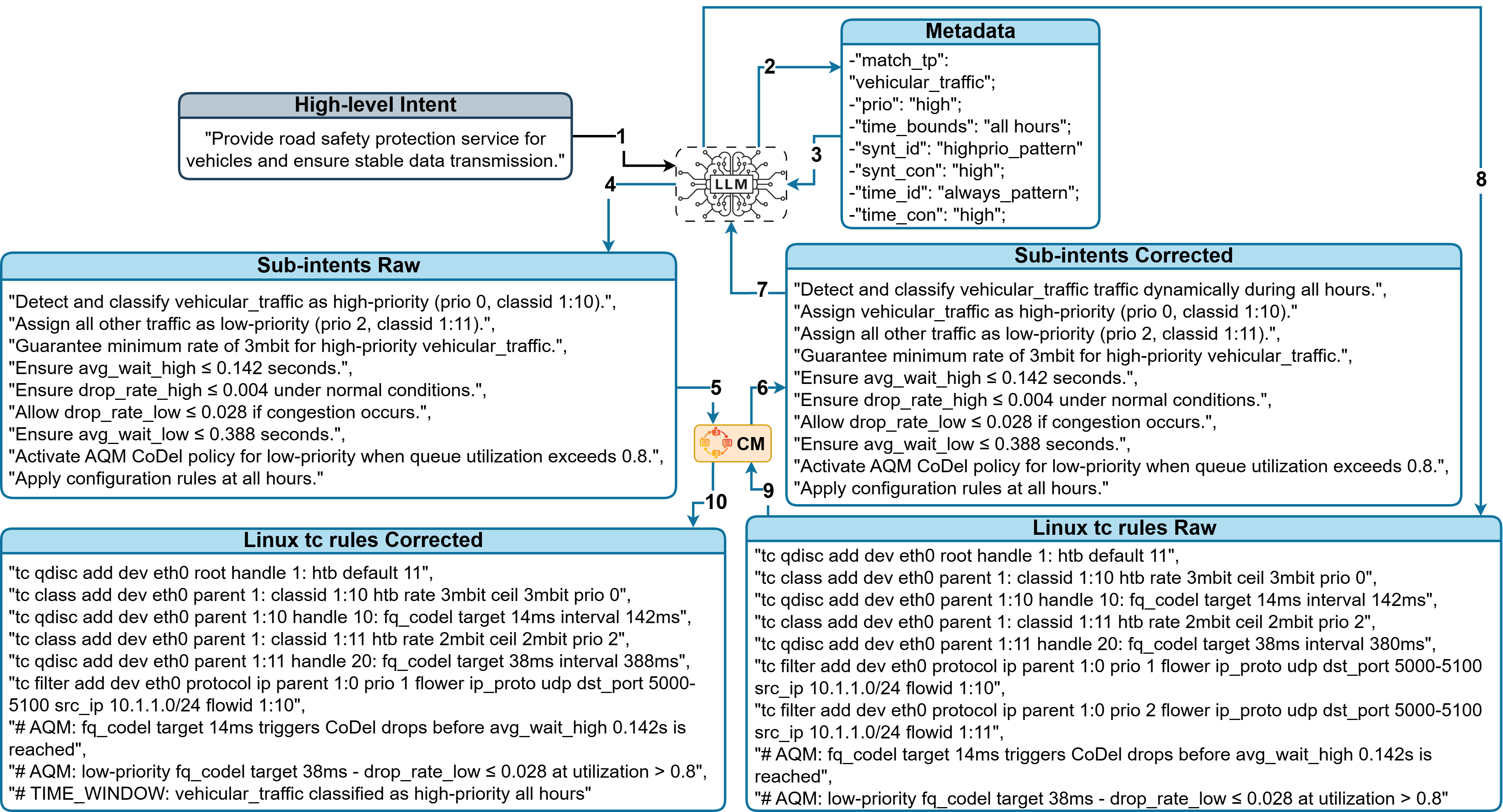}
\caption{Case study: high-level intent to critique-corrected, deployable Linux \texttt{tc} configurations; numbered arrows show processing order.}
  
  \par\smallskip\begin{minipage}{\linewidth}\raggedright\footnotesize\emph{Note:} HTB classes \texttt{1:10} and \texttt{1:11} use the \texttt{fq\_codel} AQM queueing discipline (\texttt{qdisc}); the \texttt{target}/\texttt{interval} parameters enable congestion-driven early drops and Explicit Congestion Notification (ECN). The \texttt{flower} classifier matches protocol, ports, and IP, steering traffic to high-priority \texttt{1:10} with hardware offload and no \texttt{u32} masks. Future dynamic deep packet inspection (DPI) extensions use \texttt{fwmark} for non-port flows.\end{minipage}
  \label{fig:case-study-vehicular}
\end{figure*}
\section{Performance Evaluation}
\subsection{Dataset and Evaluation Metrics}
As outlined in Section~\ref{intro}, the 100-intent benchmark was derived from the business-intent dataset of~\cite{li2025business}. We selected 100 intents spanning diverse traffic profiles, QoS objectives, and temporal requirements, and transformed them into RFC~9315-compliant traffic-shaping intents~\cite{clemm2022intent}, preserving the original operational objectives while abstracting implementation-specific parameters. Ground-truth sub-intents and Linux \texttt{tc} configurations were then derived and validated by experienced network administrators and traffic engineers, providing a deployment-ready reference for evaluating LM-generated outputs.

Fig.~\ref{fig:dataset-stats} summarizes the composition of the 100-intent dataset. As shown in Fig.~\ref{fig:qos}, the dataset covers diverse QoS objectives, including latency, bandwidth, priority, time-/load-aware control, and packet-drop constraints, with stronger emphasis on latency and priority-related goals due to their central role in practical traffic shaping. Fig.~\ref{fig:traffic} shows that the traffic taxonomy spans communication services, the Internet of Things (IoT), e-learning, emergency applications, e-commerce, video streaming, gaming, and Web traffic, ensuring coverage of heterogeneous application scenarios. 
Finally, Fig.~\ref{fig:intent} illustrates the generality of the considered intents, which include both single-constraint and multi-objective requirements, as well as always-active and time-sensitive policies.

The lack of established benchmarks for intent-to-\texttt{tc} translation motivated our dataset construction and necessitates evaluation across semantic fidelity, configuration accuracy, and deployment readiness. For sub-intent generation, we use Sentence Bidirectional Encoder Representations from Transformers (SBERT) cosine similarity~\cite{reimers2019sentence}, Recall-Oriented Understudy for Gisting Evaluation--Longest Common Subsequence (ROUGE-L) F1~\cite{lin2004rouge}, Token-F1~\cite{schutze2008introduction}, and Metric for Evaluation of Translation with Explicit ORdering (METEOR)~\cite{denkowski2014meteor}. For \texttt{tc} rule generation, we evaluate semantic unit coverage~\cite{peng2020few}, Token-F1, and normalized edit distance (NED)~\cite{yujian2007normalized}. To assess deployment readiness beyond these metrics, we incorporate the Format, Explainability, Accuracy, Cost, and Inference Time (FEACI) framework~\cite{dinh2025towards}, covering syntactic validity, technical correctness, operational feasibility, and deployment efficiency. 
Since some FEACI scores require human assessment, we derive automated counterparts (Table~\ref{tab:eval-metrics}) per the authors' definitions.

\subsection{Case Study: Stabilize Vehicular Traffic Data Transmission}
\label{sec:case-study}
To illustrate the critique-and-correction workflow, Fig.~\ref{fig:case-study-vehicular} presents a real-time vehicle-to-everything (V2X) intent. Prior to deployment, the CM performs the following corrections to the generated sub-intents and \texttt{tc} rules. In \textbf{Phase 2-B}, the CM splits a malformed sub-intent into separate profile/time-bound detection and priority-assignment sub-intents. In \textbf{Phase 2-C}, the CM removes a hallucinated filtering rule for the low-priority HTB \texttt{class} 1:11 and inserts the missing \texttt{TIME\_WINDOW} application annotation into the rule set.

\begin{table*}[!t]
\centering
\footnotesize
\resizebox{\textwidth}{!}{%
\begin{tabular}{@{}lp{0.23\linewidth}p{0.45\linewidth}@{}}
\toprule
\textbf{Metric} & \textbf{Formula} & \textbf{What it evaluates} \\
\midrule

\textbf{FEACI Format (F)} $\uparrow$ &
$\displaystyle S_F = \frac{1}{G}\sum_{j=1}^G\mathbf{1}[g_j\text{valid structurally}]$ &
Fraction of correct format: expected syntax in sub-intents and deployable \texttt{tc} rules.
\\

\textbf{FEACI Explainability (E)} $\uparrow$ &
$\displaystyle S_E = \frac{1}{G}\sum_{j=1}^G\mathbf{1}[\exists k \in K : k \subset g_j]$ &
Fraction of items with reasoning keywords or explanation comments, $K$ = set of explanation keywords.
\\

\textbf{FEACI Accuracy (A)} $\uparrow$ &
$\displaystyle S_A = \sum_{j=1}^G\frac{\sum_{m=1}^V w_m[v_m^{gold} \subset g_j]}{\sum_{m=1}^Vw_m}$ &
Weighted fraction of outputs with correct configuration values (e.g., traffic-profile match assigned higher weight).
\\

\textbf{FEACI Cost ($C_n$)} $\downarrow$ &
$\displaystyle C = c_i(N_{in})+c_o(N_{out}) \newline
S_C =
\begin{cases}
C/C_0 & \text{if } C\leq10C_0, \\
1 & \text{otherwise}
\end{cases}$ &
Normalized token cost (0 for open-source models; reference $C_0=0.1$ United States dollars (USD)).
\\

\textbf{FEACI Inference Time ($I_n$)} $\downarrow$ &
$\displaystyle S_I = \min(I/I_0,1)$ &
Normalized inference latency (threshold $I_0=60~\text{s}$).
\\

\textbf{Overall FEACI}~\cite{dinh2025towards} $\uparrow$ &
$\displaystyle FEACI = \sum_{i=F,E,A,C,I} w_i S_i$ &
Weighted FEACI composite score ($w_i=0.20$).
\\

\bottomrule
\end{tabular}}
\caption{Automated FEACI metrics adapted here for deployment readiness. \textbf{Legend:} $\uparrow/\downarrow$: higher/lower is better; $G/g_j$: generated outputs/individual output (sub-intents or \texttt{tc} rules); $V$: dataset reference values; $S_C/S_I$: normalized cost/inference scores. Established metrics:~\cite{acharya2026intent2qos,denkowski2014meteor}.} 
\label{tab:eval-metrics}
\end{table*}

\subsection{Analysis of Results}
The evaluations are conducted using static few-shot and dynamic RAG-based prompt construction, focusing on sub-intent generation from high-level intents and translation into Linux \texttt{tc} rules.

\begin{figure*}[htp]
  \centering
  \includegraphics[width=0.9\linewidth]{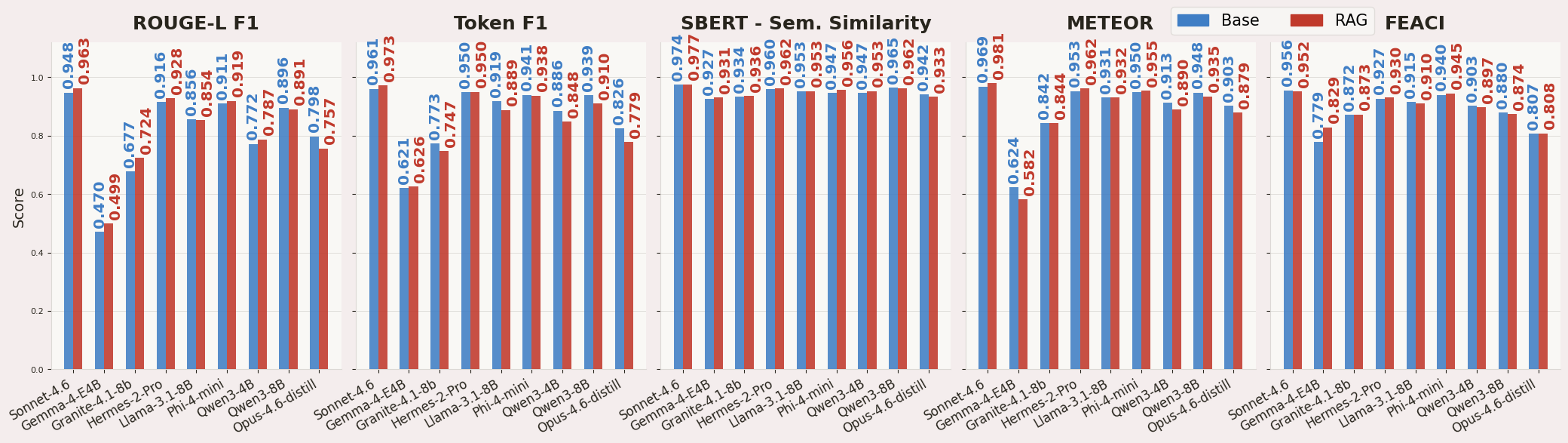}
  \caption{Evaluated LMs: five-run mean intent-to-sub-intent performance.}
  \label{fig:intent_subintent_metrics}
\end{figure*}
\begin{figure*}[htp]
  \centering
  \includegraphics[width=0.9\linewidth]{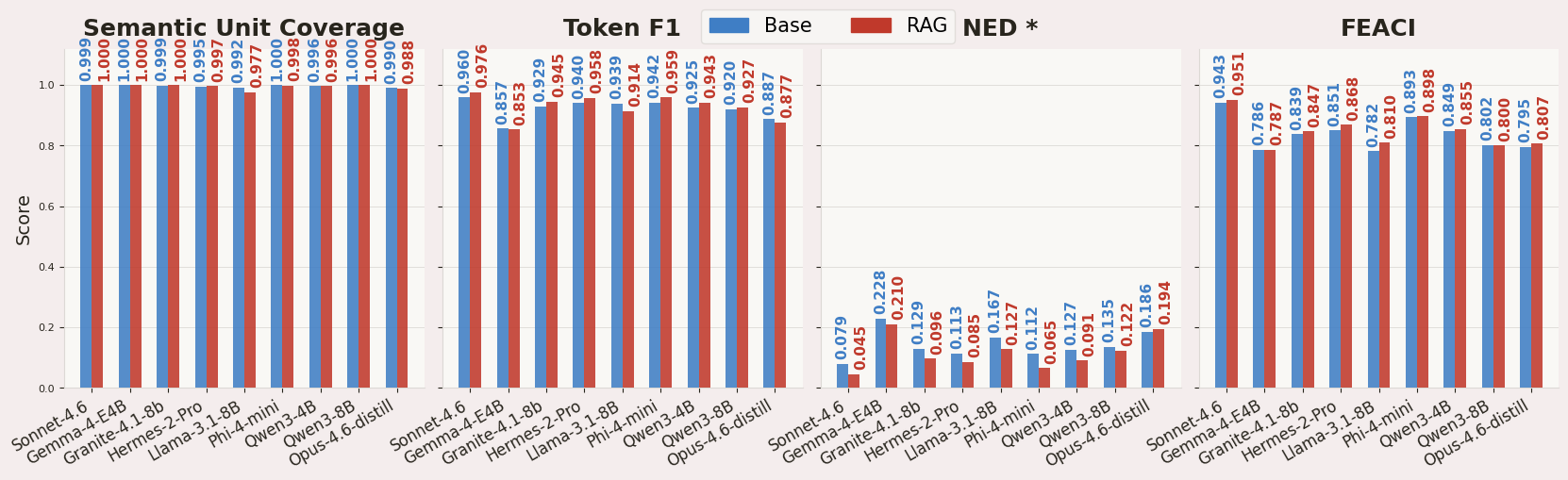}
  \caption{LMs: five-run mean sub-intent-to-\texttt{tc} performance; \footnotesize\textit{NED*}: lower is better}
  \label{fig:subintent_tc_metrics}
\end{figure*}

\begin{figure} [!htp]
    \centering
    \includegraphics[width=0.9\linewidth]{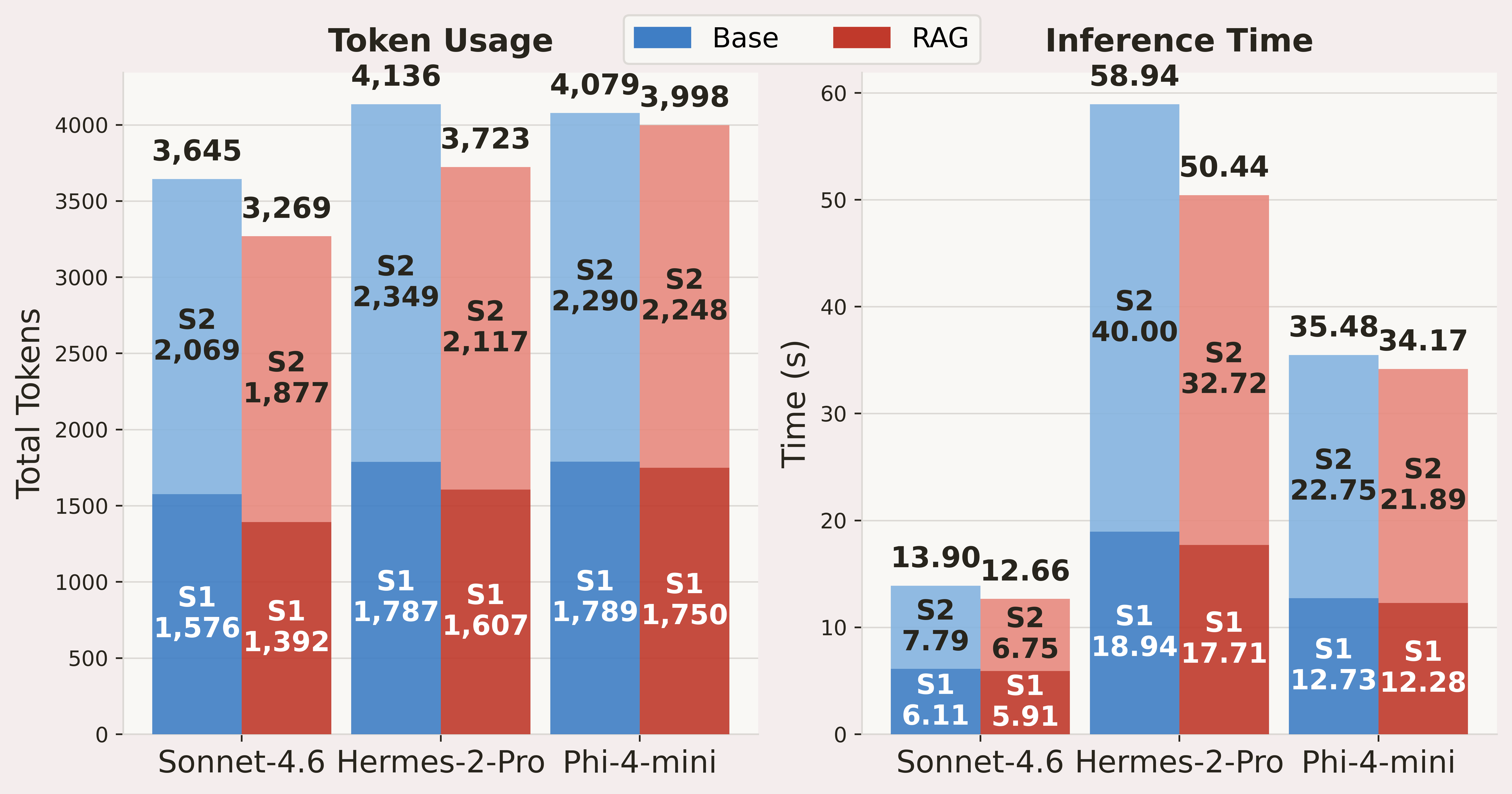}
   \caption{Top models under Base/RAG: token use and inference time; S1/S2: intent-to-sub-intent/sub-intent-to-\texttt{tc} generation.}
    \label{fig:model-usage}
\end{figure}
\subsubsection{Dynamic Metadata Extraction Analysis}
We compare the proposed LM-based dynamic metadata extraction with the static keyword-matching logic of~\cite{acharya2026intent2qos}. With Claude Sonnet-4.6 as the metadata extraction model, the dynamic approach improves traffic-profile identification accuracy from 70\% to 95\% and time-sensitive constraint detection accuracy from 59\% to 94\%, as evaluated against the ground-truth metadata of the intent dataset. This improvement reduces the dependence on manually specified keyword-to-profile mappings and enables more flexible interpretation of diverse natural-language intents, as discussed in Section~\ref{relatedWorks}.

\subsubsection{Sub-Intent Generation Results}
Fig.~\ref{fig:intent_subintent_metrics} shows that all LMs achieve strong performance across semantic, lexical, and deployment-readiness metrics, with SBERT scores generally above 0.93 and FEACI exceeding 0.80. Claude Sonnet-4.6 consistently performs best, achieving the highest scores across all metrics. Among open-source models, Hermes-2-Pro and Phi-4-mini provide the strongest overall results, indicating that the proposed framework enables even compact models to generate high-quality sub-intents.

The impact of RAG augmentation is generally modest but positive. For models such as Sonnet-4.6, Hermes-2-Pro, and Phi-4-mini, RAG consistently improves or preserves semantic fidelity and deployment-readiness scores. Gains are most evident in ROUGE-L and Token-F1, suggesting that retrieved examples better align generated sub-intents with the expected structure. Minor degradations observed for some models are mainly caused by over-specialization toward retrieved examples, resulting in verbose or merged sub-intent formulations. However, these effects remain limited, and the overall trends confirm that RAG improves consistency without compromising semantic correctness.

\subsubsection{Analysis of Generated Linux \texttt{tc} Rules}
Fig.~\ref{fig:subintent_tc_metrics} presents the Phase 2-C \texttt{tc} rule-generation results. Claude Sonnet-4.6 achieves the strongest overall performance, although SLMs such as Phi-4-mini remain highly competitive when provided with corrected sub-intents. Compared with the previous phase, performance differences across models are less pronounced, with most models achieving near-perfect semantic unit coverage and high Token-F1 scores. This reflects the deterministic nature of \texttt{tc} rules, which primarily require mapping semantic-model parameters into a rigid configuration syntax. Consequently, smaller models can closely match larger LLMs. As in the previous phase, RAG-based retrieval generally improves \texttt{tc} generation. For Sonnet-4.6, NED is nearly halved, while semantic unit coverage, Token-F1, and FEACI improve. Similar trends are observed for most models, suggesting that retrieved examples help align generated configurations with the expected rule structure. Nevertheless, even the static few-shot setup achieves strong results, highlighting the effectiveness of the critique module in producing deployment-ready \texttt{tc} configurations.

\subsubsection{Token Usage and Inference Time Analysis}
Fig.~\ref{fig:model-usage} compares the token usage and inference time of the best-performing models under Base and RAG settings. Overall, RAG reduces token consumption and latency while maintaining comparable semantic fidelity, structural precision, and deployment readiness. The gains are most evident for Claude Sonnet-4.6, where total simulation cost decreases from approximately \$1.87 to \$1.72, with average reductions of $\sim375$ tokens and $\sim1.25$ seconds per intent. Hermes-2-Pro and Phi-4-mini show similar efficiency gains, indicating that retrieved examples improve generation efficiency without sacrificing output quality.

We compare our framework with the single-pass pipeline of~\cite{acharya2026intent2qos} using Claude Sonnet-4.6, the best-performing backbone LLM in our evaluation. We observe improvements in both translation stages: in Phase 2-B, dynamic metadata extraction increases ROUGE-L, Token-F1, and SBERT by 0.07, 0.06, and 0.06, respectively; in Phase 2-C, critique-driven correction improves Token-F1 by 0.065 and reduces NED by 0.218. The proposed framework also lowers per-intent token usage by approximately 23,500 tokens, reduces inference latency by 2.2 seconds, and achieves an overall cost reduction of approximately \$6 on the full dataset.

\section{Conclusion}
This work presents a closed-loop framework for translating traffic-shaping intents into deployable Linux \texttt{tc} configurations through the integration of DT semantic modeling, critique-driven correction, and RAG-based knowledge reuse. The reported experimental results (average values across five independent runs) demonstrate high translation quality and robustness, achieving strong semantic fidelity, configuration accuracy, and deployment readiness. RAG augmentation further improves efficiency by reducing token consumption and inference latency while maintaining or improving translation quality. Notably, compact models such as Phi-4-mini achieve performance comparable to that of larger LLMs in both sub-intent and \texttt{tc} generation, while having an approximately $50\%$ smaller model size, lower random-access memory (RAM) and video random-access memory (VRAM) usage, and $33\%$ lower inference time, as shown in Figs.~\ref{fig:intent_subintent_metrics}--\ref{fig:model-usage}. 
Future work includes dynamic telemetry-driven semantic modeling, persistent multi-run RAG knowledge bases, larger-scale real-world evaluations, and CM extensions for edge-case rule conflicts.
\vspace{-0.1in}
\section*{Acknowledgment}
This work was supported in part by the Natural Sciences and Engineering Research Council of Canada (NSERC) under the DISCOVERY and CREATE TRAVERSAL Programs. 
%
%

\bibliographystyle{IEEEtran}
\bibliography{references}

@article{leivadeas2022survey,
  title={A survey on intent-based networking},
  author={Leivadeas, Aris and Falkner, Matthias},
  journal={IEEE Communications Surv. \& Tut.},
  volume={25},
  number={1},
  pages={625--655},
  year={2022},
  publisher={IEEE}
}

@techreport{clemm2022intent,
  title={{Intent-Based Networking}---Concepts and Definitions},
  author={Clemm, Alexander and Ciavaglia, Laurent and Granville, Lisandro Zambenedetti and Tantsura, Jeff},
  year={2022},
  institution={Internet Engineering Task Force},  type={Request for Comments},  number={9315},  month={October},  doi={10.17487/RFC9315},  url={https://www.rfc-editor.org/rfc/rfc9315}
}

@article{velasco2021end,
  title={End-to-end intent-based networking},
  author={Velasco, Luis and Signorelli, Marco and De Dios, Oscar Gonz{\'a}lez and Papagianni, Chrysa and Bifulco, Roberto and Olmos, Juan Jose Vegas and Pryor, Simon and Carrozzo, Gino and Schulz-Zander, Julius and Bennis, Mehdi and others},
  journal={{IEEE Communications Magazine}},
  volume={59},
  number={10},
  pages={106--112},
  year={2021},
  publisher={IEEE}
}

@inproceedings{yu2023comprehensive,
  title={A comprehensive framework for intent-based networking, standards-based and open-source},
  author={Yu, Henry and Rahimi, Hesam and Janz, Christopher and Wang, Dong and Yang, Chungang and Zhao, Yehua},
  booktitle={NOMS 2023-2023 IEEE/IFIP Network Operations and Management Symposium},
  pages={1--6},
  year={2023},
  organization={IEEE}
}

@inproceedings{dzeparoska2023llm,
  title={{LLM}-based policy generation for intent-based management of applications},
  author={Dzeparoska, Kristina and Lin, Jieyu and Tizghadam, Ali and Leon-Garcia, Alberto},
  booktitle={2023 19th International Conference on Network and Service Management (CNSM)},
  pages={1--7},
  year={2023},
  organization={IEEE}
}

@inproceedings{acharya2026intent2qos,
  title={{Intent2QoS}: Language model-driven automation of traffic shaping configurations},
  author={Acharya, Sudipta and Kantarci, Burak},
  booktitle={ICC 2026-IEEE International Conference on Communications},
  pages={1--6},
  year={2026},
  organization={IEEE}
}

@article{li2025business,
  title={Business intent and network slicing correlation dataset from data-driven perspective},
  author={Li, Jie and Zou, Sai and Sun, Yanglong and Gao, Hongfeng and Ni, Wei},
  journal={Scientific Data},
  volume={12},
  number={1},
  pages={419},
  year={2025},
  publisher={Nature Publishing Group UK London}
}

@article{pfefferle2021ieee,
  title={{IEEE} 802.1{Qcr} asynchronous traffic shaping with {Linux} traffic control},
  author={Pfefferle, Christopher and Wiedner, Florian and Schwarzenberg, Christoph},
  journal={Network},
  volume={11},
  year={2021}
}

@article{bimo2025intent,
  title={Intent-based network for {RAN} management with large language models},
  author={Bimo, Fransiscus Asisi and Galdon, Maria Amparo Canaveras and Lai, Chun-Kai and Cheng, Ray-Guang and Chong, Edwin KP},
  journal={arXiv preprint arXiv:2507.14230},
  year={2025}
}

@inproceedings{wei2025inta,
  title={{INTA}: Intent-based translation for network configuration with {LLM} agents},
  author={Wei, Yunze and Xie, Xiaohui and Hu, Tianshuo and Zuo, Yiwei and Chen, Xinyi and Chi, Kaiwen and Cui, Yong},
  booktitle={2025 IEEE 33rd International Conference on Network Protocols (ICNP)},
  pages={1--16},
  year={2025},
  organization={IEEE}
}

@article{wang2024netconfeval,
  title={{NetConfEval}: {C}an {LLMs} facilitate network configuration?},
  author={Wang, Changjie and Scazzariello, Mariano and Farshin, Alireza and Ferlin, Simone and Kosti{\'c}, Dejan and Chiesa, Marco},
  journal={Proc. ACM on Networking},
  volume={2},
  number={CoNEXT2},
  pages={1--25},
  year={2024},
  publisher={ACM New York, NY, USA}
}

@article{hossain2025netintent,
  title={{NetIntent}: Leveraging large language models for end-to-end intent-based {SDN} automation},
  author={Hossain, Md Kamrul and Aljoby, Walid},
  journal={IEEE Open Journal of the Communications Society},
  volume={6},
  pages={10512--10541},
  year={2025},
  publisher={IEEE}
}

@inproceedings{reimers2019sentence,
  title={{Sentence-BERT}: Sentence embeddings using Siamese {BERT}-networks},
  author={Reimers, Nils and Gurevych, Iryna},
  booktitle={Proc. Conf. on Empirical Methods in Natural Lang. Processing and the Intl. Joint Conf. on Natural Lang. Processing},
  pages={3982--3992},
  year={2019}
}

@inproceedings{lin2004rouge,
  title={{ROUGE}: A package for automatic evaluation of summaries},
  author={Lin, Chin-Yew},
  booktitle={Text summarization branches out},
  pages={74--81},
  year={2004}
}

@book{schutze2008introduction,
  title={Introduction to information retrieval},
  author={Sch{\"u}tze, Hinrich and Manning, Christopher D and Raghavan, Prabhakar},
  volume={39},
  year={2008},
  publisher={Cambridge University Press Cambridge}
}

@inproceedings{denkowski2014meteor,
  title={{METEOR} universal: Language specific translation evaluation for any target language},
  author={Denkowski, Michael and Lavie, Alon},
  booktitle={Proceedings of the ninth workshop on statistical machine translation},
  pages={376--380},
  year={2014}
}

@inproceedings{peng2020few,
  title={Few-shot natural language generation for task-oriented dialog},
  author={Peng, Baolin and Zhu, Chenguang and Li, Chunyuan and Li, Xiujun and Li, Jinchao and Zeng, Michael and Gao, Jianfeng},
  booktitle={Findings of the Association for Computational Linguistics: EMNLP},
  pages={172--182},
  year={2020}
}

@article{yujian2007normalized,
  title={A normalized {Levenshtein} distance metric},
  author={Yujian, Li and Bo, Liu},
  journal={{IEEE Transactions on Pattern Analysis and Machine Intelligence}},
  volume={29},
  number={6},
  pages={1091--1095},
  year={2007},
  publisher={IEEE}
}

@article{dinh2025towards,
  title={Towards End-to-End Network Intent Management with Large Language Models},
  author={Dinh, Lam and Cherrared, Sihem and Huang, Xiaofeng and Guillemin, Fabrice},
  journal={arXiv preprint arXiv:2504.13589},
  year={2025}
}

@misc{huggingface,
  author = {HuggingFace},
  title = {{Hugging Face Hub}},
  url = {https://huggingface.co},
  note = {Accessed: Jun. 2, 2026}
}

\end{document}